# Stochastic Dynamics of Bionanosystems: Multiscale Analysis and Specialized Ensembles

S. Pankavich[a], Y. Miao[b], J. Ortoleva, Z. Shreif[b], and P. Ortoleva[b]

[a]Department of Mathematics
[b]Department of Chemistry
Center for Cell and Virus Theory
Indiana University
Bloomington, IN 47405

**Abstract**
An approach for simulating bionanosystems such as viruses and ribosomes is presented. This calibration-free approach is based on an all-atom description for bionanosystems, a universal interatomic force field, and a multiscale perspective. The supramillion-atom nature of these bionanosystems prohibits the use of a direct molecular dynamics approach for phenomena like viral structural transitions or self-assembly that develop over milliseconds or longer.

A key element of these multiscale systems is the cross-talk between, and consequent strong coupling of, processes over many scales in space and time. Thus, overall nanoscale features of these systems control the relative probability of atomistic fluctuations, while the latter mediate the average forces and diffusion coefficients that induce the dynamics of these nanoscale features. This feedback loop is overlooked in typical coarse-grained methods.

We elucidate the role of interscale cross-talk and overcome bionanosystem simulation difficulties with (1) automated construction of order parameters (OPs) describing supra-nanometer scale structural features; (2) construction of OP dependent ensembles describing the statistical properties of atomistic variables that ultimately contribute to the entropies driving the dynamics of the OPs; and (3) the derivation of a rigorous equation for the stochastic dynamics of the OPs. As the OPs capture hydrodynamic modes in the host medium, "long-time tails" in the correlation functions yielding the generalized diffusion coefficients do not emerge. Since the atomic scale features of the system are treated statistically, several ensembles are constructed that reflect various experimental conditions. Attention is paid to the proper use of the Gibbs-hypothesized equivalence of long-time and ensemble averages to accommodate the varying experimental conditions. The theory provides a basis for a practical, quantitative bionanosystem modeling approach that preserves the cross-talk between the atomic and nanoscale features. A method for integrating information from nanotechnical experimental data in the derivation of equations of stochastic OP dynamics is also introduced.

**Keywords**: virus, ribosome, bionanosystems, time-of-flight experiments, chemical tagging, nanopore devices, multiscale analysis, coarse-graining, Gibbs hypothesis, Smoluchowski equation, long-time tails

# I. Introduction

Viruses, ribosomes, and other bionanosystems (BNS) are supramillion atom structures that must be understood in terms of the interplay of atomistic, highly fluctuating behaviors and more coherent dynamics involving the collective motion of many atoms simultaneously. The challenge we address is to quantify this picture and use it to understand BNS behaviors such as structural transitions, self-assembly, and interaction with other features (e.g., membranes) in a cell's interior or other biological milieu.

Being supramillion atom in size, a BNS presents a grand challenge for predictive modeling. For example, the highly optimized molecular dynamics code NAMD, run on a 1024 processor supercomputer, can simulate approximately 1.1 nanoseconds of evolution for a complete satellite tobacco mosaic virus (STMV) in one day. However, viral structural transitions take a millisecond or longer. Thus, using this approach would take about 2500 years. While some other methods taking advantage of the symmetry of the BNS (e.g., icosahedral viruses)[7] can proceed much faster, they cannot address the highly nonlinear, local and dissipative nature of the BNS phenomena of interest. Phenomenological models (e.g., wherein peptides and nucleotides are represented as beads, or other major subunits of the BNS are lumped into simplified computational elements) must be recalibrated with each new application, severely limiting their predictive power. Lumped models can also lead to difficulties in that the definition of the subunits may not be invariant over the time course of the phenomena of interest. Thus, new concepts and computational algorithms are needed for understanding and designing bionanosystems.

Advances in multiscale theory[8-10] imply an algorithm that enables all-atom dynamical computer simulations of a BNS over biologically relevant time periods. Here we present a rigorous and generalized reformulation of the all-atom multiscale approach. Multiscale analysis is a way to study systems that simultaneously involve processes on widely separated time and length scales. It has been of interest at least since the work on Brownian motion by Einstein[11-22]. In these studies, FP (Fokker-Plank) and Smoluchowski equations are derived either from the Liouville equation or via phenomenological arguments for nanoparticles without internal atomic-scale structure. We extended this work by (1) accounting for atomic-scale internal structure of the BNS; (2) introducing general sets of structural order parameters (OPs) characterizing nanoscale features of the system; (3) inventing a way to introduce OPs into the analysis without the need for tedious bookkeeping to ensure that the number of degrees of freedom is unchanged; and (4) introducing ensembles constrained to fixed values of the OPs to construct average forces and diffusion coefficients in the coarse-grained equations of stochastic dynamics[8-10]. These elements underlying our BNS modeling approach are strongly interrelated.

*Order Parameters*     Major features of a BNS change slowly in time relative to the $10^{-14}$ second characteristic timescale of fast atomic collisions and vibrations. Variables describing these nanoscale features include the CM (center-of-mass), orientation, and nanometer-scale BNS substructural units (e.g., capsomers for a virus). OPs change slowly for several reasons: (1) they have large inertia as they describe the coherent motion of many atoms simultaneously; (2) fluctuating atomic forces tend to cancel as they are averaged over the BNS surface; (3) frictional forces are large at the nanoscale and thereby repress rapid motions; and (4) energy barriers may be large relative to thermal energies. As a result, although OPs have some macroscopic character, they represent features which are small enough that many associated phenomena must be understood in terms of a stochastic description.

*Ensembles and Interscale Cross-Talk* While the OPs change slowly, the atomistic variables explore a broad range of configurations. As one is usually interested in the dynamics of the OPs, and has little concern for the detailed atomic configurations at any instant, it is relevant and practical to treat these atomistic variables using a probabilistic framework. To do so, we construct a probability density that accounts for the instantaneous values of the OPs at each point along their time course. Like total system energy for the canonical ensemble used in Gibbsian statistical mechanics of macroscopic systems, values of the OPs must be used to construct ensembles characterizing the likelihood that the system resides in each configuration of the atomistic variables. As for the thermodynamics of macroscopic systems, the construction of the ensembles developed in Section II reflect the available information on system preparation and the exchange of energy and mass during the process of interest. Unlike macroscopic systems, however, one may not always choose the ensemble that is most convenient for carrying out the calculations, that is, the finite size of the system requires greater attention when constructing the ensemble and choosing among several possible ensembles, as noted in Sect. II.

*Perturbation Theory and Coarse-Grained Stochastic Order Parameter Dynamics* One may identify small parameters (notably mass, time, length ratios, and force strengths) that can be used to develop an orderly perturbation solution of the Liouville equation for the *N*-atom probability density. The lowest order solution is constructed via an OP-constrained entropy maximization principle and a closely related restatement of the Gibbs hypothesized equivalence of ensemble and long-time averages. Expansion methods based on the smallness of these ratios imply equations for the reduced probability density describing the stochastic evolution of the OPs. Depending on the character of the system and the physical/biological regime of interest, the stochastic equation derived is of the Smoluchowski or Fokker-Planck type. The frictional and average force factors appearing in these equations depend sensitively on the values of the OPs and on the conditions (e.g., iso-energetic or isothermal) to which the system is subjected. Our multiscale perspective provides insight into the cross-talk among the atomistic and nanoscale variables. As suggested in **Fig. 1**, the OPs set the overall (nanoscale) context that determines the relative probability of atomic configurations, which in turn, mediate the average forces and frictional effects that drive the dynamics of the OPs. Such a relationship is the basis of the feedback loop expressed in **Fig. 1**. This is in sharp contrast to typical coarse-grained models, which aim to simplify the system by reducing the number of interacting elements. This is usually done via "decoupled" coarse-graining wherein atoms are lumped together, and an effective force on the lumped element is set forth[1-6]. In particular, a systematic method to derive the parameters needed for the effective force field from trajectories and force data collected from a single MD simulation was developed[3]. This approach was applied to various systems such as liquids[4], monosaccharides[5], and peptides[6]. The results were in good agreement with atomistic simulations while computational cost was significantly reduced. Note however, that MD simulations can only be carried over short time and length scales. Thus, one wonders if the derived parameters would still be applicable over longer scales. Furthermore, the derived parameters are used as a basis of the coarse-grained simulation without regard for the feedback loop of **Fig. 1** (i.e., the role of the overall structure on the computation of the coarse-grained forces is overlooked). For example, the ensemble of fluctuations of atoms deep within a tightly wound globular protein is likely to differ significantly from that when the protein is uncoiled. A similar comment is appropriate for the transition state between two distinct conformations separated by a high energy barrier. A further difficulty is that this type of coarse-graining provides no self-consistency criterion regarding the number of atoms to be used in a lumped element. The ensemble of atomic-scale fluctuations that should be used in a proper computation of the coarse-grained forces depends on temperature and other conditions, thus the parameters in the coarse-graining should change with these conditions. Trajectories produced in this manner are only reliable over short times wherein the overall nanostructure doesn't change enough to significantly alter the ensemble of atomistic fluctuations.

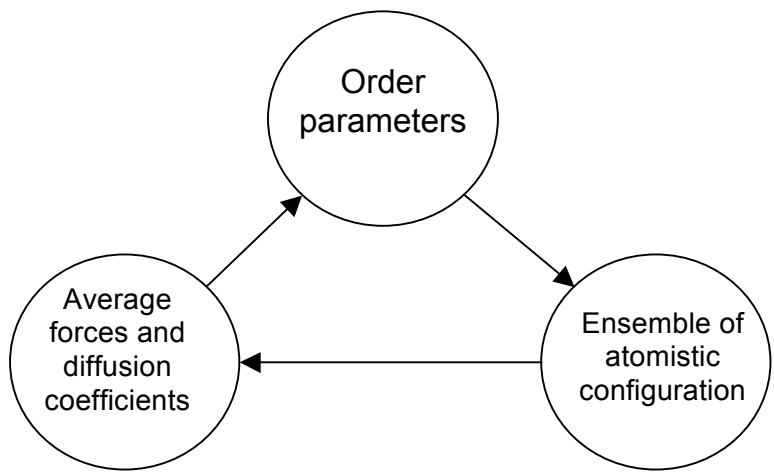

**Fig. 1** OPs characterizing nanoscale features affect the relative probability of the atomistic configurations which, in turn mediates the forces driving OP dynamics. This feedback loop is central to a complete multiscale understanding of nanosystems and the true nature of their dynamics.

The feedback loop of **Fig. 1** suggests the need for the simulation algorithm in **Fig. 2**. The coarse-grained equations for stochastic OP dynamics are solved numerically. As the average forces and diffusion coefficients in these equations change with the OPs, they must be co-evolved with the OPs. These factors are expressed as statistical averages weighted with OP-constrained probability densities.

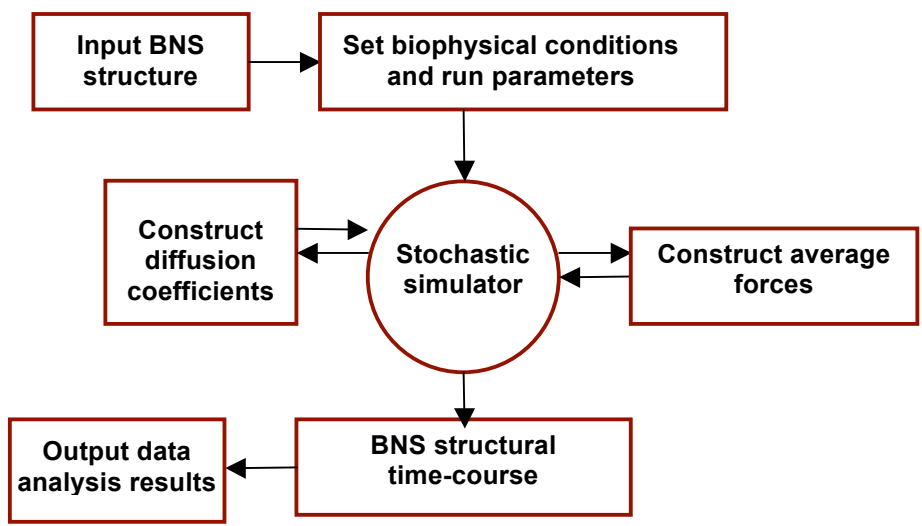

**Fig. 2** In the fully coupled multiscale simulator, the stochastic equations of the Langevin type for the BNS structural OP dynamics are simulated. Diffusion coefficients and average forces are computed on-the-fly as they change with the OPs via the feedback loop of Fig. 1.

In Section II, we construct the probability distribution for the atomistic variables via an entropy-maximization principle and the use of OPs as constraints. In Section III, we introduce a procedure for automatically constructing BNS OPs for viruses, ribosomes, membranes, and other bionanostructures. In Section IV, we show how the construction of appropriate ensembles and multiscaling can be interwoven into a BNS analysis. Conclusions are drawn in Section V.

**II. Ensembles for Atomistic Fluctuations of a BNS**

A BNS is envisioned to have dual macroscopic/microscopic characteristics - nanoscale components, each consisting of many atoms, and atomic-scale, highly fluctuating features. To characterize the coherent nature of the nanoscale features, OPs are introduced and expressions relating them to the all-atom configuration are developed (see Section III and appendix). These OPs transcend particular atomic associations; for example, the CM of a nanoparticle is a collective property of thousands of atoms, not of a single atom. Involving so many atomic notions in a coherent way, OPs in a BNS typically change slowly in time relative to the timescale of individual atomic vibrations and collisions (e.g., $10^{-14}$ to $10^{-12}$ seconds). Thus, associated with the coherent nanometer-scale, stochastic atomic-scale dichotomy, there is a corresponding timescale separation. The objective of the formulation developed here is to simultaneously account for these diverse scales and the cross-talk among them.

A common theme of many statistical mechanical theories of nanosystems is that the timescale separation suggests that the atomistic variables explore a representative set of configurations over a period of time on which the OPs are relatively constant. The relative residence time in each member of this ensemble of configurations is characterized by a probability density $\rho$. By definition, $\rho$ quantifies the likelihood that the short-scale behaviors visit each configuration in this ensemble. Given our lack of knowledge of most of the atomic-scale detail, and that most of it is not directly relevant to understand BNS behavior, stability, and function, an entropy-maximization approach is used to construct $\rho$. We develop this approach and integrate it with multiscale analysis in this and subsequent sections. In the statistical mechanics of macroscopic systems, this enables the study of various experimental conditions (e.g., isothermal versus iso-energetic). Here, we show that this can be used to introduce nanotechnical measurement information into the analysis. Given knowledge of some nanoscale features of the system derived from nanotechnical measurements, our objective is to construct the probability distribution for the rapidly fluctuating degrees of freedom.

The statistical state of a BNS for most biologically relevant conditions evolves slowly, i.e., with a characteristic time much greater than that ($10^{-12}$ to $10^{-14}$ seconds) of atomic vibrations and collisions. In Section IV, we show that this implies that the probability density $\rho$ characterizing this statistical state must depend on the $\Gamma$ (the set of 6$N$ atomic positions and momenta) only via the set $\Phi$ $(=\{\Phi_1, \cdots \Phi_{M+1}\})$ of $M$ structural OPs plus the energy $(\Phi_{M+1})$. In contrast, if $\rho$ depends on $\Gamma$ in other ways, then $\rho$ evolves on the atomic time scale. Thus, while the lowest order analysis of the $N$-atom system implies that $\rho$ has the restricted form $\rho(\Phi)$, it does not provide further information on the specific functional form of $\rho$. Since no information about the detailed atomic-scale state of the system is usually known, we turn to an information theoretic method to construct $\rho$.

According to Jaynes[23], the entropy $S$ for the classical system as formulated here for the nanosystem is given in terms of an integral of $\rho \ln \rho$. For the restricted state counting as in Appendix A, this becomes

$$S = -k_B \int \omega d\Gamma^* \Delta(\varphi - \Phi(\Gamma^*)) \rho \ln \rho \qquad \textbf{(II.1)}$$

where $\Delta$ is a product of $M+1$ Dirac delta functions, one for each of the M structural OPs and one for the energy. The construction of $\rho$ follows from a consideration of the particular experiment under investigation. Several distinct cases are analyzed below.

*Iso-energetic Closed Systems* Consider an isolated system in which the energy is held constant at a value $E$. To construct $\rho$, we maximize $S$ subject to normalization

$$\int \omega d\Gamma^* \Delta(\varphi - \Phi(\Gamma^*)) \rho = 1. \qquad \textbf{(II.2)}$$

Since energy is automatically fixed to $\varphi_{M+1}$ $(=E)$ via the $\delta$-function mediated state counting, no further constraints on $S$ maximization are required here. Using the Lagrange multiplier approach, one obtains the probability density denoted in the present case by $\hat{\rho}_\varphi$ in the form

$$\hat{\rho}_\varphi = 1/Z(\varphi) \qquad \textbf{(II.3)}$$

$$Z(\varphi) = \int \omega d\Gamma^* \Delta(\varphi - \Phi(\Gamma^*)). \qquad \textbf{(II.4)}$$

Using (**II.1**) and (**II.2**) one finds that $Z = e^{S/k_B}$. Since gradients of $Z$ with respect to the OPs will be shown in Section IV to derive coherent OP dynamics, one concludes that in this iso-OP ensemble, evolution is driven by entropy differences. This is an expression of the Second Law for an isolated nanosystem.

We consider $\hat{\rho}_\varphi$ to be a conditioned probability density. If $W_\varphi d^{M+1}\underline{\varphi}$ is the probability that the system is in a state where the $M+1$ OPs are in a small element $d^{M+1}\varphi$ about the particular value $\varphi$ (i.e., $\int d^{M+1}\varphi W_\varphi = 1$), then the corresponding probability density $\rho_\varphi$ is $\hat{\rho}_\varphi W_\varphi$. Here and in the following, we drop the $d^{M+1}\varphi$ for simplicity as all the equations for $\rho$ (as in Section IV) are linear so that such factors are found to cancel in the equations. With the above, the overall probability $\rho_\varphi$ for this case is given by

$$\rho_\varphi = \hat{\rho}_\varphi W_\varphi, \qquad \textbf{(II.5)}$$

a general form that, with modification of the $\hat{\rho}$ and $W$ factors, holds for all cases considered in this study.

A superficial examination of (**II.4**) suggests that $Z$ is not bounded. However, integration over the $3N$ atomic positions is bounded by the volume of the vessel containing the system. Furthermore, the delta function on energy restricts the momentum integrations because the total energy is a constant, the potential energy is bounded from below (i.e., the potential energy can never approach $-\infty$ as the core atom-atom potential becomes $+\infty$ as two atoms overlap), and the energy is a monotonically increasing function of all the atomic momenta.

*Isothermal Closed Systems* Consider an ensemble of systems in contact with a thermal bath so that the energy $H(\Gamma)$ is only known by its average $\langle H \rangle$,

$$\langle H \rangle = \int \omega d\Gamma^* \Delta(\underline{\varphi} - \underline{\Phi}(\Gamma^*)) H(\Gamma^*) \rho, \qquad \textbf{(II.6)}$$

where $\underline{\Phi}$ is the set of $M$ structural OPs, $\underline{\varphi}$ is a particular set of their values, and $\omega = \omega_0 \varsigma_1 \cdots \varsigma_M$ (see Appendix A). With this average energy constraint and that of normalization, entropy maximization implies

$$\rho_{\underline{\varphi}\beta} = \frac{e^{-\beta H(\Gamma)}}{Q(\underline{\varphi},\beta)} W_{\underline{\varphi}\beta} \equiv \hat{\rho}_{\underline{\varphi}\beta} W_{\underline{\varphi}\beta} \tag{II.7}$$

$$Q(\underline{\varphi},\beta) \equiv \int \omega d\Gamma^* \Delta(\underline{\varphi} - \underline{\Phi}(\Gamma^*)) e^{-\beta H(\Gamma^*)}. \tag{II.8}$$

The isothermal partition function $Q$ is related to the free energy $\hat{F}_{\underline{\varphi}\beta} \equiv \langle H \rangle_{\underline{\varphi}\beta} - \hat{S}_{\underline{\varphi}\beta}/k_B \beta$ via $Q = e^{-\beta \hat{F}_{\underline{\varphi}\beta}}$. This result was obtained in an earlier study[9] in a more circuitous manner.

*Mixed Energy Systems* In two stage experimental protocol, the system is initially equilibrated with a thermal bath, and allowed to evolve in an energy-conserving vessel. For example, a nanoparticle first traverses a thermalizing gas and is then redirected to a vaccine chamber for iso-energetic evolution. The statistical analysis for such an experiment proceeds in two related stages. First, one solves the problem iso-energetically arriving at $\hat{\rho}_{\underline{\varphi}E}$ for particular energy value $E$. Taking the energy reference state such that $H \geq 0$, the statistical average of any quantity $A(\Gamma^*)$ in this mixed ensemble takes the form

$$\langle A \rangle_{\text{mixed}} = \beta \int_0^\infty dE e^{-\beta E} \int d^M \varphi \int \omega d\Gamma^* \Delta(\underline{\varphi} - \underline{\Phi}(\Gamma^*)) \delta(E - H(\Gamma^*)) \hat{\rho}_{\underline{\varphi}E} W_{\underline{\varphi}E} A(\Gamma^*), \tag{II.9}$$

i.e., the problem is solved for each particular energy $E$ and the result is then averaged via the canonical weight $\beta e^{-\beta E}$ and $W_{\underline{\varphi}E}$.

*Nanotechnical Experiments* The ensembles constructed via the entropy maximization procedure as above are designed to experimentally integrate derived information with the theory. For example, the classic canonical ensemble imposes the temperature to construct the ensemble of atomistic configurations while the microcanonical ensemble imposes the energy to arrive at the ensemble. In a similar manner, we treat experimental approaches wherein the apparatus is designed to constrain the system so that the uncertainty in the spectrum of atomistic fluctuations is decreased. As the allowed spectrum of atomistic configurations and fluctuations determines the average forces and diffusion coefficients determining the stochastic OP dynamics, this experimental information can greatly improve our capability to predict BNS dynamics.

*1. Fluorescence Tagged Ensembles*
Fluorescence or mass labeling can be used to identify parts of a BNS which reside near its surface during the process of interest[24]. The label is presented via the microenvironment and only penetrates within a short distance below the BNS surface. The notion of a molecular surface has previously been quantitatively defined[25]. This definition provides an algorithm for constructing a surface denoted here $\Sigma(\Gamma, \vec{r})$ such that if the atomic configuration is $\Gamma$, then points $\vec{r}$ for which $\Sigma < 0$ lie within the structure and those for which $\Sigma > 0$ are in the microenvironment. In our tagged ensemble approach, we first construct the average depth below the $\Sigma = 0$ surface where the atoms in the labeled parts of the BNS reside. For arbitrary $\Gamma$, $d$ will be on the order of magnitude of the size of the BNS. By experimental design, it is known that $d$ is small for configurations during the process of interest. Thus, to eliminate irrelevant configurations, we limit the count of states by adding a factor $\theta(d_c - d)$ to all integrals over $\Gamma^*$, where $d_c$ is a depth to which the labeling compound penetrates below the BNS surface. The tagging of near-surface components of a BNS yields information about the system; thus, we expect to have a related decrease in entropy $S$,

$$S = -k_B \int \omega d\Gamma^* \Delta(\varphi - \Phi(\Gamma^*)) \theta(d_c - d(\Gamma^*)) \rho \ln \rho. \quad \textbf{(II.10)}$$

The additional information has the effect of decreasing the entropy and thereby altering the driving force for OP dynamics. By construction, $S$ is an increasing function of $d_c$. Thus, if the tagging is only very close to the surface and these tagged BNS subunits are always near the surface, then any configuration $\Gamma$ with these subunits far below the BNS surface would have a low probability and thereby be avoided by the stochastic OP dynamics we derive (see Section IV). Though the BNS remains with the tagged subunits near the surface in one experiment, they may not remain so in another. Thus, a simulation that showed a divergence of the tagged entropy descending to large negative values because of energetic driving forces would indicate a structural inversion (i.e., inside-out) transition.

*2. The TOF Ensemble*
Time-of-flight experiments measure the cross-section for collision between the background gas molecules and the BNS. Efficient algorithms exist for computing $\sigma(\Gamma)$, the cross-section for the specific microconfiguration $\Gamma$ of the BNS[26,27]. Thus, if $\sigma_o$ is the observed value, then to the counting factor we add $D(\sigma - \sigma_o, \zeta_{c-s})$, which is unity when $\sigma$ is within $\pm \zeta_{c-s}$ of $\sigma_o$ and zero otherwise. For example, when the OPs cause $D$ to approach zero, the TOF entropy approaches $-\infty$. As the average force on the OPs is the gradient of the TOF entropy with respect to the OPs, this implies that the dynamics of the TOF ensemble member systems be driven to evolve away from such states. As shown in Section IV, the Langevin equations we develop will guarantee that if the OPs are initially consistent with $|\sigma - \sigma_o| < \zeta_{c-s}$, they will remain so for the simulated time-course.

*3. The Nanopore Ensemble*
Confined BNS evolution in a nanopore is being studied for several reasons[28]: (1) the BNS is fixed in space so that it can be subjected to electrical, dielectric, or other fields and mechanical disturbances, and (2) the size of the pore can be chosen to only allow the entry of nanoparticles in a given size range. Furthermore, once the BNS is totally fitted into a nanopore, zero electric forces can be induced in the pore material and forces thereby applied to induce structural transitions. The presence of the nanopore is already accounted for in our formulation as it is reflected in the limits on the positions of the atoms – i.e., they remain in the system which is within the nanopore.

A goal of multiscale analysis is to construct the coarse-grained probability density $\tilde{W}_\varphi$. For an experiment represented by $\hat{\rho}_\varphi W_\varphi$ one obtains

$$\tilde{W}_\varphi = \int d\varphi' \int \omega d\Gamma^* \Delta(\varphi' - \varphi) \rho_{\varphi'}(\Gamma^*) W_{\varphi'} = W_\varphi. \quad \textbf{(II.11)}$$

This states that $W$ is the reduced probability density for $\varphi$ the $N$-atom distribution as constructed here. However, as the multiscale analysis develops, we find that the $\hat{\rho} W$ form does not capture corrections due to the slow evolution of the OPs described in the Liouville equation.

In our earlier treatment[9], we also constructed $\rho$, but in two steps. We present that approach in Appendix B to contrast that presentation to the present one.

To proceed with the multiscale analysis, we develop OPs suited for a spectrum of BNS (Section III) and then use them with a multiscale analysis of the Liouville equation to determine the time dependence of the statistical state of these systems (Section IV). The properties of the ensembles constructed affect the character of the stochastic OP dynamics. One such property is the average $\langle \vec{p}_i \rangle$ of the momentum of a single atom (*i* here). The probability density in all the above cases depends on $\vec{p}_i$ via a $p_i^2/2m_i$ term in the kinetic energy. This term is symmetric (even) in $\vec{p}_i$ while $\vec{p}_i$ itself is asymmetric (odd). Thus, $\langle \vec{p}_i \rangle$ involves the integral over a symmetric range in each component of $\vec{p}_i$ (i.e., $\int_{-a}^{a} dp_{ix}$ for the *x*-component of $\vec{p}_i$) with $a = \infty$ for the isothermal case and $a = \sqrt{2m_i E}$ for an isolated system at energy *E*. This implies $\langle \vec{p}_i \rangle = \vec{0}$. Thus, the average of any linear combination of the $\vec{p}_i$ is zero, a conclusion shown in Section IV to affect the form of the stochastic equation for OP dynamics.

### III. OPs for Bionanosystems

A key element of multiscale analysis of a BNS is the identification of OPs that describe its nanoscale features. The multiscale analysis requires that these parameters must have several characteristics. They should be flexible, meaning, applicable to a system of multiple nanoscale components such as viruses, functionalized nanoparticles, or cell membranes. They must capture the hydrodynamic modes of the host fluid to avoid "long-time tail" difficulties in the construction of diffusion coefficients[16] that arise in the theory of Brownian motion. The set of OPs must be "complete", i.e., they do not couple to other slowly evolving variables omitted from the set. For practical reasons, OPs should also be implementable in an automated computational procedure. In contrast, one might divide the system into "structural units" (e.g., protein complexes or viral capsomers). Unfortunately, such structural units are often artificial, i.e., are more geometric than a natural consequence of the interatomic forces and the level of thermal agitation. Finally, a central characteristic of an OP is that it evolves slowly relative to the timescale of atomic vibrations and collision. In a sense these parameters define the multiscale character of the *N*-atom system. In Section IV we show that failure to base a multiscale analysis on such OPs is at the heart of the weakness of other coarse-grained approaches.

The first studies to observe such behaviors were conducted by both Rahman[29] and by Alder and Wainwright[30,31] using molecular dynamics. Rahman determined the velocity autocorrelation function in liquid argon, while Alder and Wainwright showed that the velocity autocorrelation function of a Brownian particle in a hard-disc and hard-sphere fluid decays as $t^{-d/2}$, where *d* is the dimensionality. Long time tails were explained by a hydrodynamic model wherein the movement of the Brownian particle creates a vortex in the host fluid which in turn affects the behavior of the Brownian particle. This constitutes an additional delayed effective interaction between the particle and its host medium. Hence, hydrodynamic modes in the host medium create memory effects and long-time tails in the correlation function which can cause anomalies in the diffusion coefficients.

Since the work of Alder and Wainright, there have been several attempts to derive the asymptotic time behavior of the velocity autocorrelation function[32-34]. This was also shown experimentally using light-scattering[35,36], diffusive wave spectroscopy[37-39], and neutron scattering[32]. For historical record, however, it is important to mention that the first hydrodynamic theory of translational Brownian motion was provided by Vladimirsky and Terletzky[41] in 1945. These authors derived an equation that is equivalent to that used later for the mean square displacement of the Brownian particle[42].

To start our multiscale analysis and to include OPs that capture hydrodynamics and other slow modes, we set forth a methodology based on a deformation of space[43] that our recent studies suggest is ideally suited even for complex BNS phenomena[44]. A central property of these or other OPs is that they evolve slowly. Slow OP dynamics emerge in several ways including:
- by inertia associated with the coherent dynamics of many atoms evolving simultaneously;
- in the case of migration over long distances;
- via stochastic forces that tend to cancel;
- in species population levels as in chemical kinetics or self-assembly which involve many units, only a few of which change in an interval of time relative to the collision time.

These factors are realized for a number of variables:
- structural parameters characterizing objects such as viral capsids or ribosomes that stay intact during a transition;
- orientational angles for a nanostructure;
- density-like variables characterizing the hydrodynamic modes or composition;
- scaled positional variables describing the motion of disconnected molecules across a nanostructure such as an oil droplet;
- curvilinear/twist parameters describing the conformation of RNA, DNA and proteins[45].

While the OPs and factors that make them evolve slowly require approaches differing in technical details, the methods presented here can be extended to them all.

To illustrate our methodology, we focus on intact nanostructures. OPs for this case are introduced by embedding the system in a volume $V_s$. Basis functions $U_{\underline{k}}(\vec{s})$ for a triplet of labeling indices $\underline{k}$ are introduced which are orthonormalized. If computations are carried out using periodic boundary conditions to simulate a large system (e.g., to minimize boundary effects and to handle Coulomb forces), periodic basis functions can be used. Here, the nanosystem deforms in 3-D space. Points $\vec{s}$ in this space are considered to be a displacement of original points $\vec{s}^0$. As presented in our earlier study[36], a set of vector OPs $\vec{\Phi}_{\underline{k}}$ are constructed using orthonormal polynomials in atomic coordinates via the procedure of Appendix C:

$$\sum_{\underline{k}} B_{\underline{q}\underline{k}} \vec{\Phi}_{\underline{k}} = \sum_{i=1}^{N} m_i U_{\underline{q}}(\vec{s}_i^0) \vec{s}_i \ , \ B_{\underline{q}\underline{k}} = \sum_{i=1}^{N} m_i U_{\underline{q}}(\vec{s}_i^0) U_{\underline{k}}(\vec{s}_i^0) \qquad \textbf{(III.1)}$$

where $m_i$ is the mass of atom $i$ ($i = 1, 2, \cdots N$) and the integral orthogonality of the basis functions implies that the matrix $B_{\underline{q}\underline{k}}$ is nearly diagonal.

With these OPs, a multiscale Molecular Dynamics/Order Parameter extrapolation (MD/OPX) approach has been developed to simulate large BNS[36]. In the approach, a short MD run estimates the rate of change of the OPs, which is then used to extrapolate the system over times that are much longer than the $10^{-14}$ second timescale of fast atomic vibrations and collisions. It has been shown that the OPs satisfy all criteria for a multiscale computational approach and the simulator has provided one to two orders of magnitude computational speed-up over direct MD.

## IV. Multiscale Theory of BNS Kinetics

An equation of stochastic OP dynamics is now obtained via a multiscale analysis for a classical $N$-atom system that preserves the feedback between the atomistic and nanoscale variables of **Fig. 1**. The rapidly fluctuating degrees of freedom are hypothesized to explore a representative sample of configurations during a fraction of one characteristic time for the dynamics of the OPs. Using the Gibbs hypothesis, an average over such a time interval is equated to an average over the representative sample of configurations of the rapidly fluctuating variables. In Section II such an ensemble is constructed to be consistent with given values of the slowly changing OPs. Thus, such ensembles are a key element of our multiscale BNS theory. In this section we use these concepts and the Liouville equation to derive equations for the slow stochastic dynamics of a BNS.

The analysis starts by writing the Liouville equation for the $N$-atom probability density $\rho$, i.e., $\partial \rho / \partial t = \mathcal{L} \rho$ for Liouville operator $\mathcal{L}$. Many authors (see [3, 4] for reviews) have recast this equation in the form $\partial \rho / \partial t = (\mathcal{L}_0 + \varepsilon \mathcal{L}_1) \rho$ for small parameter $\varepsilon$ (e.g., a mass ratio as in Section III). This recast equation is solved perturbatively via a Taylor expansion in $\varepsilon$.

Recently[10,11,45,46], we have argued that the above multiscale form of the Liouville equation arises by making the ansatz that $\rho$ depends on the $N$-atom state $\Gamma$ both directly, and indirectly via a set of OPs, $\Phi$. In adopting this perspective, $\Phi$ is not a set of additional independent dynamical variables, rather its appearance in $\rho$ is a place-holder for a special dependence of $\rho$ on $\Gamma$ that underlie the slow temporal dynamics of $\rho$. This perspective avoids the need for tedious bookkeeping wherein a number of atomistic variables are removed to preserve the total number (**6N**) of degrees of freedom. The dual dependence of $\rho$ on $\Gamma$ can be ensured if $\varepsilon$ is sufficiently small. In turn, this is assured if $\Phi$ is slowly varying in time relative to the timescale ($10^{-14}$ to $10^{-12}$ seconds) of atomistic fluctuations.

Because many BNS processes depend sensitively on their internal atomistic structure, and earlier studies on nanoparticles via the Liouville equation ignored this internal structure, we develop an approach that is symmetric with respect to all atoms in the system (e.g., those in a virus, cell membrane, and aqueous medium). In the following, we derive an equation for the OP probability distribution in two cases: (1) an isolated system of given energy and (2) an isothermal system. The mixed and other cases of Section II are addressed briefly.

*A. The Isolated System at Energy E*

The coarse-grained probability density $\tilde{W}_{\underline{\varphi} E}$ for a system of energy $E$ with structural OPs $\underline{\varphi}$ is defined via

$$\tilde{W}_{\underline{\varphi} E} = \int \omega d\Gamma^* \Delta\big(\underline{\varphi} - \underline{\Phi}(\Gamma^*)\big) \delta\big(E - H(\Gamma^*)\big) \rho . \qquad \textbf{(IV.1)}$$

The operators $\mathcal{L}_0$ and $\mathcal{L}_1$ noted above arise out of the ansatz on the dual dependence of $\rho$ on $\Gamma$ (i.e., $\rho(\Gamma^*, \Phi, t)$) and the chain rule. Thus, $\mathcal{L}_0$ involves partial derivatives with respect to $\Gamma$ at constant $\Phi$ (when operating on $\rho$ in the multiscale form $\rho(\Gamma, \Phi, t)$), and conversely for $\mathcal{L}_1$. Again, this does not imply that the $\Phi$ are dynamical independent variables. Rather, this is a reflection of the dual dependence of $\rho$ $(=\rho(\Gamma, \Phi, t))$. However, as the system is isolated, $\mathcal{L}_1$ has no derivative with respect to $\Phi_{M+1} = H$. This occurs as the walls of the vessel are perfectly reflecting, that is, conserving of kinetic energy and redirecting velocity.

As shown in the analysis of the OPs (Section III), the parameter $\varepsilon$ ($<<1$) naturally emerges from Netwon's equations, i.e., when computing $d\Phi_k/dt = -\mathcal{L}\Phi_k \equiv \varepsilon \Pi_k$ for momentum-like variable $\Pi_k(\Gamma)$. Here we use it to organize the construction of $\rho$. In Section III, $\varepsilon$ was found to be the ratio of small-to-large characteristic masses (e.g., the mass of a typical atom to that of the major nanoscale components of a BNS). For simplicity of notation, we label the OPs such that $k = 1, 2, \cdots M$ are the structural OPs of Section III and $k = M+1$ indicates the energy $H$.

As with other multiscale approaches reviewed in Section I, it is postulated that $\rho$ depends on the sequence of times $t_0, t_1, t_2, \cdots = t_0, \underline{t}$ where $t_n = \varepsilon^n t$. The times $t_n$ for $n > 0$ are introduced to account for the slower behaviors in $\rho$, while $t_0$ accounts for processes on the fast timescale (i.e., $t_0$ changes by one unit when $10^{-14}$ seconds elapse).

With the above framework, the Liouville operators $\mathcal{L}_0$ and $\mathcal{L}_1$ take the form

$$\mathcal{L}_0 = -\sum_{i=1}^{N} \frac{\vec{p}_i}{\vec{m}_i} \cdot \frac{\partial}{\partial \vec{r}_i} + \vec{F}_i \cdot \frac{\partial}{\partial \vec{p}_i} \quad \text{(IV.2)}$$

$$\mathcal{L}_1 = -\underline{\Pi} \cdot \frac{\partial}{\partial \underline{\Phi}}, \quad \text{(IV.3)}$$

where $\underline{\Phi}$ is the set of all OPs except energy $H$. There is no $\partial/\partial H$ term in $\mathcal{L}_1$ since the conjugate momentum $\Pi_{M+1}$ for $H$ is zero due to the iso-energetic nature of the walls continuing the system. Note, $m_i, \vec{p}_i, \vec{r}_i$ and $\vec{F}_i$ are the mass, momentum, position and net force for atom $i$. Since neither $\mathcal{L}_0$ nor $\mathcal{L}_1$ involves a derivative with respect to $H$, $\rho$ only depends on energy parenthetically. If $E$ is the fixed value of $H$ over the evolution in the iso-energetic ensemble of interest here, then the probability density is denoted $\rho_E(\Gamma, \underline{\Phi}, t_0, \underline{t})$. The strategy we now pursue is to construct $\rho_E$ via an asymptotic expansion in $\varepsilon$ and use it to construct an equation for the evolution of $\tilde{W}_{\underline{\varphi} E}$.

The Gibbs hypothesized equivalence of long-time and ensemble averages plays a key role in our multiscale analysis of the Liouville equation. The lowest order Liouville operator $\mathcal{L}_0$ has a corresponding propagator $e^{-\mathcal{L}_0 t_0}$ that evolves dynamical variables (notably any function of $\Gamma$) in time, yet leaves $\underline{\Phi}$ unchanged. Thus, the Gibbs hypothesis takes the form

$$\lim_{t \to \infty} \frac{1}{t} \int_{-t}^{0} dt' e^{-\mathcal{L}_0 t'} A = \langle A \rangle_{\underline{\varphi} E}. \quad \text{(IV.4)}$$

Here $\langle \cdots \rangle_{\underline{\varphi} E}$ indicates an average of any variable over all configurations selected by the factor $\Delta(\underline{\varphi} - \underline{\Phi}^*)$ and weighted by the conditional probability $1/Z(\underline{\varphi})$ for $\varphi = \underline{\varphi}, E$, i.e.,

$$\langle A \rangle_{\underline{\varphi} E} = \int \omega d\Gamma^* \Delta(\underline{\varphi} - \underline{\Phi}(\Gamma^*)) \delta(E - H(\Gamma^*)) A(\Gamma^*) / Z(\underline{\varphi}, E). \quad \text{(IV.5)}$$

This provides key elements needed to carry out a multiscale analysis for iso-energetic systems.

We seek a perturbative solution of the Liouville equation in the form $\rho = \sum_{n=0}^{\infty} \varepsilon^n \rho_n$ with quasi-equilibrium character to lowest order, i.e., $\rho_0$ is independent of $t_0$ and arises form entropy maximization. To lowest order in $\varepsilon$ we find

$$\mathcal{L}_0 \rho_0 = 0 \,. \tag{IV.6}$$

Recalling results from Section II for the iso-energetic case and that any function of $\Phi$ is in the nullspace of $\mathcal{L}_0$, one obtains

$$\rho_0 = \hat{\rho}_{\underline{\varphi}E} W_{\underline{\varphi}E} \tag{IV.7}$$

for $\hat{\rho}_{\underline{\varphi}E} = 1/Z(\underline{\varphi}, E)$ and factor $W_{\underline{\varphi}E}$ to be determined.

To $O(\varepsilon)$ the multiscale Liouville equation implies

$$\left(\frac{\partial}{\partial t_0} - \mathcal{L}_0\right)\rho_1 = -\left(\frac{\partial}{\partial t_1} - \mathcal{L}_1\right)\rho_0 \,. \tag{IV.8}$$

Taking $\rho_1$ to be $A_1$ at $t_0 = 0$, this equation has the solution

$$\rho_1 = e^{\mathcal{L}_0 t_0} A_1 - t_0 \hat{\rho} \frac{\partial W_{\underline{\varphi}E}}{\partial t_1} - \int_{-t_0}^{0} dt_0' e^{-\mathcal{L}_0 t_0'} \underline{\Pi} \cdot \frac{\partial}{\partial \underline{\Phi}}\left(\frac{W_{\underline{\Phi}E}}{Z(\underline{\Phi}, E)}\right). \tag{IV.9}$$

Since the system is bounded in space by the walls of the vessel, and as the potential energy approaches $+\infty$ when any two atoms overlap, $e^{-\mathcal{L}_0 t_0} A_1$ for any function $A_1$ of $\Gamma$ and $\Phi$ fluctuates but remains finite for all $t_0$. With this, there is no term to balance the divergent $t_0$ contribution to $\rho_1$ and hence $\partial W_{\underline{\varphi}E} / \partial t_1$ must vanish.

A general equation for the coarse-grained probability density $\tilde{W}$ for any of the ensembles of Section II can be obtained by using the Liouville equation. By definition, $\tilde{W}_{\underline{\varphi}E}$ is related to the *N*-atom probability density $\rho$ via (**IV.1**). The Liouville equation implies

$$\frac{\partial \tilde{W}_{\underline{\varphi}E}}{\partial t} = \int \omega d\Gamma^* \Delta(\varphi - \Phi(\Gamma^*)) \mathcal{L} \rho \tag{IV.10}$$

for *N*-atom density $\rho$. Properties of the delta function $\Delta$, $\mathcal{L}H = 0$, and integration by parts imply

$$\frac{\partial \tilde{W}_{\underline{\varphi}E}}{\partial t} = -\varepsilon \frac{\partial}{\partial \underline{\varphi}} \cdot \int \omega d\Gamma^* \Delta(\varphi - \Phi(\Gamma^*)) \underline{\Pi} \rho \,. \tag{IV.11}$$

Developing $\rho$ in a series in $\varepsilon$, one can construct the (*n*+1)-th correction to the rate of change of $\tilde{W}_{\underline{\varphi}E}$ from the *n*-th order correction to $\rho$. For the OPs of Section III and the triple index $\underline{k}$ labeling used there, one obtains

$$\vec{\Pi}_{\underline{k}} = \frac{1}{\hat{m}} \sum_i U_{\underline{k}}(\vec{s}_i^{\,0}) \vec{p}_i \,. \tag{IV.12}$$

As noted in Section II, the ensemble average of any of the individual atomic momenta is zero, therefore so is the iso-energetic ensemble average of $\underline{\Pi}$. The factor $W_{\underline{\varphi}E}$ in $\rho_0$, the lowest order solution for $\rho$, is seen to be the lowest order contribution to $\tilde{W}_{\underline{\varphi}E}$ in $\varepsilon$. With the above, we see that the $O(\varepsilon)$ contribution to $\partial \tilde{W}_{\underline{\varphi}E} / \partial t$ is zero, and hence $\partial \tilde{W}_{\underline{\varphi}E} / \partial t$ must be $O(\varepsilon^2)$.

Using the expression for $\rho_1$ and the definition of $\tilde{W}_{\underline{\varphi}E}$, it is found that the only $O(\varepsilon)$ contribution to $\tilde{W}_{\underline{\varphi}E}$ is from $e^{\mathcal{L}_0 t_0} A_1$. Notice that $A_1$, the initial $(t_0 = 0)$ data for $\rho_1$, is arbitrary, i.e., depends on the experiment of interest. Thus, for some initial data, it is seen that the $O(\varepsilon)$ contribution to $\tilde{W}_{\varphi E}$ can have short timescale dependence (e.g., due to a shock wave). However, for most BNS phenomena we do not expect such phenomena to be part of the experimental design. This implies that for the BNS phenomena of interest, $A_1$ is in the null space of $\mathcal{L}_0$, i.e., depends on $\Gamma$ only through $\Phi$. The result is that in order for **(IV.11)** to be closed in $\tilde{W}_{\varphi E}$ to $O(\varepsilon^2)$, $A_1$ must be a function of $\tilde{W}_{\varphi E}$ only. For the special case $A_1 = 0$, $\tilde{W}_{\varphi E} = W_{\varphi E}$ up to $O(\varepsilon^2)$, and this is the special case studied henceforth, i.e., we assume the system is initially in the $\hat{\rho}_{\varphi E} W_{\varphi E}$ quasi-equilibrium state.

Collecting the above results for the special case $A_1 = 0$ with iso-energetic conditions implies a Smoluchowski equation for $\tilde{W}_{\varphi E}$:

$$\frac{\partial \tilde{W}_{\varphi E}}{\partial t} = -\varepsilon^2 \sum_{k=1}^{M} \frac{\partial}{\partial \varphi_k} \bullet J_{k\varphi E} \tag{IV.13}$$

$$J_{k\varphi E} = -\sum_{l=1}^{M} D_{kl\varphi E}\left[\frac{\partial}{\partial \vec{\varphi}_l} - \langle f_l \rangle_{\varphi E}\right] \tilde{W}_{\varphi E}, \ k = 1, 2, \cdots M \tag{IV.14}$$

$$D_{kl\varphi E} = \frac{1}{m^2} \int_{-\infty}^{0} dt' \langle \Pi_k e^{-\mathcal{L}_0 t'} \Pi_l \rangle_{\varphi E}. \tag{IV.15}$$

As $\hat{\rho}_{\varphi E} = 1/Z(\underline{\varphi}, E)$ and $Z = e^{\hat{s}/k_B}$ (see Section II), it is seen that $k_B \langle \vec{f}_k \rangle = \partial \hat{S}/\partial \varphi_k$ for the closed iso-energetic system. Hence, OP dynamics is entropy-driven. This is a natural result for an isolated system.

The above analysis can be extended to higher order in $\varepsilon$. If all initial data appearing in higher-order corrections to $\rho$ (i.e., the $A_2 A_3, \cdots$) are zero, closure of the equation for $\tilde{W}_{\varphi E}$ is ensured. This does not create difficulties in resolving secular behavior in the higher order analysis as discussed in detail elsewhere[47].

Corresponding to the Smoluchowski equation derived above there is an equivalent set of friction-dominant (i.e., non-inertial) Langevin equations for the stochastic dynamics of $\Phi$. These equations provide a convenient way to simulate a BNS via an algorithm as in **Fig. 2**. As $\langle f_k \rangle_{\varphi E}$ and $D_{ll'\varphi E}$ depend on $\underline{\varphi}$ they must be computed at each timestep. Thus, the feedback in **Fig. 1** is accounted for.

*B. Isothermal Systems*

The development for a system at temperature $1/k_B \beta$ closely follows that of Section IV.A except that the lowest order solution $\rho_0$ is given by

$$\rho_0 = \frac{e^{-\beta H}}{Q(\underline{\Phi}, \beta)} W_{\Phi\beta}. \tag{IV.16}$$

It must be recognized that this is a heuristic treatment. For example, a system in contact with a thermal bath experiences continuous exchange of energy through the bounding walls so that in a more rigorous treatment the Liouville equation must be solved with stochastic boundary conditions.

Again, the objective is to derive an equation for the probability density, which for this case is given by

$$\tilde{W}_{\underline{\varphi}\beta} = \int \omega d\Gamma^* \Delta(\underline{\varphi} - \underline{\Phi}(\Gamma^*))\rho_\beta, \quad \textbf{(IV.17)}$$

for isothermal $N$-atom probability density $\rho_\beta$. One proceeds in a similar manner as above but with subtle differences. Using the expression for $\hat{\rho}_{\underline{\varphi}\beta}$, we find $\langle \bar{f}_k \rangle_{\underline{\varphi}\beta} = -\partial \hat{F}_{\underline{\varphi}\beta}/\partial \varphi_k$ for free energy $\hat{F}_{\underline{\varphi}\beta}$ see Section II.B. As before, we find a Smoluchowski equation for $\tilde{W}_{\underline{\varphi}\beta}$ up to $O(\varepsilon^2)$:

$$\frac{\partial \tilde{W}_{\underline{\varphi}\beta}}{\partial t} = -\varepsilon^2 \sum_{k=1}^{M} \frac{\partial}{\partial \varphi_k} \bullet \underline{J}_{\underline{\varphi}\beta k}, \quad \textbf{(IV.18)}$$

$$J_{k\underline{\varphi}\beta} = -\sum_{l=1}^{M} D_{kl\underline{\varphi}\beta} \left[ \frac{\partial}{\partial \varphi_l} - \langle f_l \rangle_{\underline{\varphi}\beta} \right] \tilde{W}_{\underline{\varphi}\beta}, \quad k=1,2,\cdots M,, \quad \textbf{(IV.19)}$$

$$D_{kl\underline{\varphi}\beta} = \frac{1}{m^2} \int_{-\infty}^{0} dt' \langle \Pi_k e^{-\mathcal{L}_0 t'} \Pi_l \rangle_{\underline{\varphi}\beta}, \quad k,l=1,2,\cdots M. \quad \textbf{(IV.20)}$$

This result is similar to that for the iso-energetic case except that the diffusion coefficients and average forces are modified to reflect the ensemble of atomistic fluctuations for the $\varphi\beta$ versus the $\varphi E$ ensemble. Unlike for the thermodynamic limit $N \to \infty$, these factors could be quite different for the two ensembles.

*C. Initial Isothermal, Iso-energetic Evolution Cases*

Consider the two stage experiment wherein the system is first equilibrated with a thermal bath and subsequently allowed to evolve iso-energetically. Since the Liouville equation is linear, a superposition of solutions also satisfies the equation. Thus, the solution corresponding to a linear combination of states with different initial probability densities, each of which is thereafter iso-energetic, can be summed with a weighting by the probability that the system had a given energy initially. As noted in Section II, the probability density for the energy of systems equilibrated with a bath at temperature $(k_B \beta)^{-1}$ is $\beta e^{-\beta e} E$ (for the energy conversion $E \geq 0$). Therefore

$$\tilde{W}_{\underline{\varphi}\text{mix}}(\underline{\varphi},t) = \beta \int_0^\infty dE e^{-\beta E} \int \omega d\Gamma^* \Delta(\underline{\varphi} - \underline{\Phi}(\Gamma^*)) \delta(E - H(\Gamma^*)) \rho(\Gamma^*,t) = \beta \int_0^\infty dE e^{-\beta E} \tilde{W}_{\underline{\varphi}E}. \quad \textbf{(IV.21)}$$

where subscript mix indicates the two stage experiment. This result shows that $\tilde{W}_{\underline{\varphi}\text{mix}}$ and $\tilde{W}_{\underline{\varphi}E}$ are related via Laplace transformation. However, the average forces and diffusion coefficients for the iso-energetic case cannot simply be averaged via the $\beta e^{-\beta E}$ weight and used in a Smoluchowski equation for $\tilde{W}_{\underline{\varphi}\text{mix}}$. Rather, these factors must be used in the Smoluchowski equation for $\tilde{W}_{\underline{\varphi}E}$ and the solution transformed to obtain $\tilde{W}_{\underline{\varphi}\text{mix}}$.

*D. Nanotechnical Experiments*

Ensembles were considered in Subsection II.D for various nanoscale experiments. They were (1) chemical labeling to distinguish subunits of a nanostructure on its outer surface; (2)

confinement of a BNS in a nanopore; and (3) selecting nanoparticles of a given cross-section area throughout the process of interest. Each case has a distinct partition function and thereby average force $\langle \vec{f}_k \rangle$ and diffusion coefficients $D_{kl}$ that mediate the evolution of the OPs. The Langevin equations for the three cases can be used to study structural transitions of a BNS under the given conditions. Ensembles of the various types could reveal particular aspects of the system. For example, a labeling experiment could reveal results on swelling transitions in a viral capsid since the penetration depth of the labeling reflects the diffusibility of the label molecule in the swollen versus normal state.

Inclusion of step function factors in the state counting of Section II allows for a situation wherein there is a range of $\varphi$ values for which the partition function is zero. This implies that as this region of $\underline{\varphi}$–space is approached, $\langle \vec{f}_k \rangle$ will diverge, driving the system away from this region. This is how the step function condition in the expression for the partition function guides the Langevin evolution away from configurations prohibited by the experimental design (e.g., a virus cannot swell beyond the volume of a nanopore in which it is trapped.

**V. Conclusions**
A methodology to obtain Smoluchowski equations for the stochastic dynamics of OPs characterizing key nanoscale features of a BNS is presented. Unlike decoupled coarse-graining methods, the key feedback of **Fig. 1** accounting for the modification of the average forces on the instantaneous values of OPs is accounted for in our fully coupled multiscaling strategy. Our development begins with the automated construction of OPs $\Phi$ that are demonstrated via Newton's equations to evolve on timescales long relative to that of atomistic fluctuations. Probability densities are developed to characterize the statistics of rapidly fluctuating atomistic degrees of freedom. The densities explore a representative set of configurations during a fraction of the characteristic time on which the OPs evolve. The atomistic degrees of freedom and the OPs are captured by the dependence of the aforementioned probability density on the OPs and, conversely, the contribution of the rapidly fluctuating to the entropy and the determination of OP dynamics by averaged energy. In this sense, our methodology can be termed fully coupled multiscaling. This is in contrast to decoupled coarse-graining where the effect of the evolving OPs on the computation of the forces on lumped elements is ignored.

In our approach, the Liouville equation is solved perturbatively upon identification of a smallness parameter $\varepsilon$, that is a ratio of characteristic masses, lengths, times, or energies. The result is a Smoluchowski equation for the stochastic dynamics of the OPs. The average force and diffusion coefficients that appear in this equation depend on the OPs. Contrastingly, in a decoupled method, coarse-grained forces are computed without regard to the OP-dependent averaging that is always changing as, for example, a viral structural transition unfolds. A reconsideration of ensembles of atomistic fluctuations (Section II) allows one to tailor the Gibbs hypothesis to a variety of special cases that introduce modern nanotechnical experimental approaches. These include fluorescent or mass labeling, TOF selected nanosystems, and nanopore confined experiments. Our use of the Gibbs hypothesis, rather than integration over atomistic degrees of freedom as in other multiscale approaches, enables full theory/experiment integration.

As the systems of interest are nanometer in scale, the usual equivalence of results of different ensembles (e.g., the microcanonical and canonical) may not hold; rather, these differences may reflect distinct experimental protocols, and alter the resulting average forces and diffusion coefficients that appear in the Smoluchowski equation.

Our multiscale analysis suggests that such a Smoluchowski equation can only be derived when the perturbation parameter is sufficiently small. Thus, coarse-graining is not meaningful unless the lumped variables for which forces are computed evolve significantly slower than the atomistic ones. Thus, coarse-grained potentials for amino acids or nucleotides may not provide a viable starting point as they are still rapidly fluctuating variables. Conversely, models where, for example, a whole pentamer or hexamer is considered to be a lumped element, may not be viable either (i.e., the internal dynamics of these entities may be a key element of self-assembly as they deform during assembly). Furthermore, these internal degrees of freedom provide a sink for friction-associated energy transfer.

The Smoluchowski equations we obtain are correct to $O(\varepsilon^2)$ even though the Liouville equation need only be solved to $O(\varepsilon)$. Thus, our procedure is ideal for deriving higher order (augmented) Smoluchowski equations even for complex systems involving multiple OPs. This feature of our workflow follows from the use of the general equation for the reduced probability density $\tilde{W}$ that follows directly from the Liouville equation.

These results enable the algorithm for BNS simulations suggested in **Fig. 2**. The Smoluchowski equation is solved in a Monte Carlo fashion via the equivalent Langevin equation with diffusion coefficients and averaged forces evolving with the OPs. While the MD timestep must be shorter than the $10^{-14}$ second timescale of fast atomic vibrations and collisions, those for the present algorithm are limited by the characteristic timescale of the phenomena of interest (e.g., $10^{-3}$ seconds for viral structural transitions). The resulting algorithm can therefore be many orders of magnitude faster than MD, despite the overhead from the computation of average forces and diffusion coefficients.

The computation of diffusion coefficients must be carried out by constructing a time correlation function with evolution generated by the operator $\mathcal{L}_0$, and not by Newton's equations directly. Thus, one may not use available MD codes to construct them. However, as the OPs are slowly varying, standard MD codes can be used to compute the diffusion coefficients approximately as long as the MD code is only run for a time short compared to the characteristic time of OP evolution, i.e., the constancy of the OPs must be checked during the construction of the diffusions coefficients.

**Acknowledgements**

This project at the Center of Cell and Virus Theory was supported by the U.S. Air Force, the U.S. Department of Energy, the Lilly Endowment, Inc., the Oak Ridge Institute for Science and Education, and the National Institute of Health.

**Appendix A: State Counting**

The first step in ensemble construction as in Section II is to develop a way to estimate the number of distinct quantum states in the 6$N$ dimensional volume element of atomic position/momentum space. For example, there is one state in a volume $N!\hbar^{3N}$ for $N$ identical atoms. If $\Delta\left(\varphi\text{-}\Phi(\Gamma^*)\right)$ is a product of $M$+1 Dirac delta functions ($M$ for the structural OPs and one for the system energy), then the aforementioned state count is given by

$$\int \omega d\Gamma^* \Delta\left(\varphi\text{-}\Phi(\Gamma^*)\right) = \begin{cases} \text{number of quantum states available to the system} \\ \text{with } \Phi \text{ in a fixed, narrow range about } \varphi, \end{cases} \quad \textbf{(A.1)}$$

where $\Gamma^* = \{\vec{p}_1, \vec{r}_1, \cdots \vec{p}_N, \vec{r}_N\}$ specifies the state of the $N$-atom system over which integration is taken. The state density factor $\omega$ is constructed as follows. Let $\theta(x)$ be the unit step function: $\theta = 1$ for $x > 0$ and $\theta = 0$ for $x \leq 0$. Then $D(x,\varsigma) = \theta(x+\varsigma/2) - \theta(x-\varsigma/2)$ is 1 for $-\varsigma/2 < x < \varsigma/2$ and zero otherwise. If $\varsigma$ is small, then $\theta(x+\varsigma/2) \approx \theta(x) + (d\theta/dx)\varsigma/2$. But $d\theta(x)/dx = \delta(x)$ for Dirac delta function $\delta(x)$. With this, $\theta(x+\varsigma/2) - \theta(x-\varsigma/2) \approx \varsigma \delta(x)$. The factor $D(\varphi_1 - \Phi_1, \varsigma_1) \cdots D(\varphi_{M+1} - \Phi_{M+1}, \varsigma_{M+1})$ is zero except in a small zone of $\Phi$ values around $\varphi$ of volume $\varsigma_1 \cdots \varsigma_{M+1}$ for width $\varsigma_k$ in each of the OPs, $k = 1, 2, \cdots M+1$. Let $1/\omega_0$ be the product of the $N!\hbar^{3N}$ state counting factors for each of the types of atoms in the system. Then, the number of quantum states in the zone for which the product of $D$ factors is one, is given by $\int \omega_0 d\Gamma^* D(\varphi_1 - \Phi_1, \varsigma_1) \cdots D(\varphi_{M+1} - \Phi_{M+1}, \varsigma_{M+1})$. Comparing this result with (**A.1**) for small $\varsigma_k$ implies $\omega = \omega_0 \varsigma_1 \cdots \varsigma_{M+1}$. The $\varsigma_k$ are all small constant quantities. Thus, $\omega$ is a constant, i.e., independent of $\Gamma^*$. With this counting of states for the iso-OP systems, in Section II we make information theoretic arguments to construct $\rho$ for a range of distinct experimental conditions relevant for the study of BNS as follows.

**Appendix B: Fourier Transform Method**

In our earlier study, $S$ was first maximized with respect to $\rho$, constrained by normalization and the ensemble average values of a set of OPs $\Phi$. In the present notation, this becomes

$$\int \omega d\Gamma^* \Phi(\Gamma^*) \rho \text{ fixed.} \quad \textbf{(B.1)}$$

Introduce Lagrange multipliers $K_1, K_2, \cdots K_{M+1}$ associated with the $M+1$ constraints (**B.1**) for the $M+1$ OPs constrained entropy maximization. One obtains the conditional probability $\hat{\rho}_K$ given by

$$\rho_K = \frac{e^{-K \cdot \Phi}}{\Xi(K)} \quad \textbf{(B.2)}$$

$$\Xi = \int \omega d\Gamma^* e^{-K \cdot \Phi(\Gamma^*)}.$$

For the second level, we develop a more general solution in the form of a linear combination of the $\hat{\rho}_K$ with unit-normalized weight $\Psi(K)$; in analogy with our earlier study[9], we find that the more general distribution, denoted $\rho_\varphi$, takes the form

$$\rho_\varphi = \frac{\Omega}{N(\varphi)} \tag{B.3}$$

$$N(\varphi) = \int d\Gamma^* \Delta(\varphi - \Phi(\Gamma^*)) \tag{B.4}$$

for ($M$+1)-fold delta function $\Delta$ and partition function and coarse-grained probability $\Omega$ $\left(\int d^{M+1}\Phi\Omega = 1\right)$. The distribution (**B.3**) is microcanonical in character for the detailed atomic configurations consisting of all states with fixed value $\varphi$ of the OPs; thus, it has an equi-a priori principle character, i.e., all states with a given $\varphi$ are equally likely in the absence of additional information.

The result (**B.3**) is derived from (**B.2**) using properties of the Fourier transform. Consider a function $f(x)$ of a set $x$ of $L$ variables. The transform $\hat{f}(k)$ for the set of $L$ "wave vectors" $k$ is defined via

$$\hat{f}(k) = \int d^L x e^{k \cdot x} f(x). \tag{B.5}$$

The inverse relation reads

$$f(x) = \frac{1}{(2\pi)^L} \int d^L k e^{-ik \cdot x} \hat{f}(ik), \tag{B.6}$$

where $i = \sqrt{-1}$. For functions $f_1(x)$ and $f_2(x)$ one has

$$\frac{1}{(2\pi)^L} \int_{-\infty}^{+\infty} d^L k e^{-ik \cdot x} \hat{f}_1(ik) \hat{f}_2(ik) = \int dx' f_1(x - x') f_2(x'). \tag{B.7}$$

The result (**B.7**) follows from the inversion formula which yields

$$\frac{\Psi(K)}{\Xi(K)} = \frac{1}{(2\pi)^{M+1}} \int d^{M+1}\varphi e^{-iK \cdot \varphi} \frac{\Omega(i\varphi)}{N(i\varphi)}. \tag{B.8}$$

Thus, $\Psi/\Xi$ is related to the integration of $\Omega/N$ over all imaginary OP values.

As $\Psi$ is arbitrary, further assumptions related to the nature of $\Omega$ that stem from the problem of interest must be introduced, e.g., the initial data. Finally, the thermal case with $\hat{\rho}_{\Phi,\beta} = e^{-\beta H}/Q(\underline{\Phi},\beta)$ corresponds to keeping the energy $(\Phi_{M+1} = H)$ fixed only by its average so that $K_{M+1} = -\beta$.

**Appendix C: Constructing order parameters from orthonormal polynomials**

Consider a nanostructure embedded in a volume $V_S$. Basis functions $U_{\underline{k}}(\bar{s})$ for a triplet of labeling indices $\underline{k}$ are introduced which are orthonormalized. In this method, the nanosystem deforms in 3-D space, which is considered to be a displacement of an original point $\vec{s}^{\,0}$. Deformation of space taking any $\vec{s}^{\,0}$ to $\vec{s}$ is continuous and is used to introduce OPs $\vec{\Phi}_k$ via

$$\vec{s} = \sum_k U_k\left(\vec{s}^{\,0}\right)\vec{\Phi}_k \,. \tag{C.1}$$

As the $\vec{\Phi}_k$ change, $\vec{s}$ space is deformed, and so does the nanosystem embedded in it. The objective is to ensure that the dynamics of the $\vec{\Phi}_k$ reflects the physics of the BNS and that the deformation reflects key aspects of the atomic-scale details of the structure. In this way, the $\vec{\Phi}_k$ constitute a set of vector OPs that serve as the starting point of our multiscale approach.

In our approach, atom $i$ ($i = 1, 2, \cdots N$) is moved from its original position $\vec{s}_i^{\,o}$ via the above deformation by evolving the $\vec{\Phi}_k$ and correcting for atomic-scale details as follows. Given a finite truncation of the $k$ sum in (C.1), there will be some residual displacement for individual atoms. Denoting this residual for atom $i$ as $\vec{\sigma}_i$,

$$\vec{s}_i = \sum_k \vec{\Phi}_k U_k\left(\vec{s}_i^{\,0}\right) + \vec{\sigma}_i \,. \tag{C.2}$$

The size of $\vec{\sigma}_i$ can be minimized by the choice of basis functions and the number of terms in the $k$ sum. Conversely, imposing a permissible size threshold for the residuals allows one to determine the number of terms to include in the sum.

To start the multiscale analysis, the $\vec{\Phi}_k$ must be expressed in terms of the fundamental variables $\vec{s}_i$. Let $m_i$ be the mass of atom $i$. Multiplying (C.2) by $m_i U_q\left(\vec{s}_i^{\,0}\right)$ and summing over the $N$ atoms in the system, one obtains

$$\sum_k B_{qk}\vec{\Phi}_k = \sum_{i=1}^N m_i U_q\left(\vec{s}_i^{\,0}\right)\vec{s}_i \,, \quad B_{qk} = \sum_{i=1}^N m_i U_q\left(\vec{s}_i^{\,0}\right)U_k\left(\vec{s}_i^{\,0}\right) \tag{C.3}$$

The integral orthogonality of the basis functions implies that the matrix $B_{qk}$ is nearly diagonal. Thus, the OPs can easily be computed in terms of the atomic positions by solving (C.3) numerically. More specifically, when most of the system is occupied with atoms, the $i$ sum is essentially a Monte Carlo integration. The orthonormality of the basis functions implies that $B_{qk} \approx \delta_{qk}$ and (C.3) can be approximated as

$$\vec{\Phi}_k \approx \frac{V_s}{N}\sum_{i=1}^N \frac{m_i}{m} U_k\left(\vec{s}_i^{\,0}\right)\vec{s}_i \,. \tag{C.4}$$

The $\vec{\sigma}_i$ contribution is neglected in arriving at this *definition* of $\vec{\Phi}_k$ as $\vec{\sigma}_i$ fluctuates in direction and magnitude with $i$, while the basis functions that capture overall characteristics of the nanosystem (e.g., position, orientation, nanoscale structure, and hydrodynamic modes) vary smoothly by design. Thus, (C.3) is not an approximation; rather, the above is a way to argue for the *definition* of OPs that express coherent behaviors of the nanosystem. With this definition of the $\vec{\Phi}_k$, (C.3) is an exact relationship, since the $\vec{\sigma}_i$ correct errors in the displaced atomic positions over-and-above the coherent contribution from the $\vec{\Phi}_k$ sum.

Our approach has several conceptual and technical advantages. In the above we include all atoms in the system, e.g., those in the nanostructure and its microenvironment. This captures the boundary layer of water that tends to accompany a nanoparticle due to the viscous nature of water at the nanoscale and the interaction forces. Thus, the method captures hydrodynamic modes in the host fluid and layers of water bound to the nanoparticles and other structures.

Inclusion of $m_i$ in the above expressions gives $\vec{\Phi}_{\underline{k}}$ the character of generalized, center-of-mass (CM) variables. For example, if $U_{\underline{k}}$ is a constant for a specific $\underline{k}$, the corresponding value of $\vec{\Phi}_{\underline{k}}$ is proportional to the CM of the system. As $d\vec{s}_i/dt = \vec{p}_i/m_i$ for momentum $\vec{p}_i$ of atom $i$, one has

$$\frac{d\vec{\Phi}_{\underline{k}}}{dt} = \frac{\vec{\Pi}_{\underline{k}}}{Nm},$$

$$\vec{\Pi}_{\underline{k}} = V_s \sum_{i=1}^{N} U_{\underline{k}}(\vec{s}_i^{\,0})\vec{p}_i.$$

(C.5)

While $\vec{\Phi}_{\underline{k}}$ has a sum of $N$ atoms, many of which have similar directions due to the smooth variation of $U_{\underline{k}}$ with respect to $\vec{s}_i^{\,0}$, the momenta have fluctuating direction and tend to cancel near equilibrium. Hence the thermal average of $\vec{\Pi}_{\underline{k}}$ is small, $\vec{\Phi}_{\underline{k}}$ tend to evolve slowly, and the ratio of the characteristic time of $\vec{\Phi}_{\underline{k}}$ to that of atomic vibrations and collisions should be on the order of the number of atoms in the system, i.e., O($N$). This suggests that the $\vec{\Phi}_{\underline{k}}$ are slowly varying, and therefore satisfies a key criterion to be an order parameter and serve as the starting point of our multiscale analysis. As $N$ is large, it is convenient to define the smallness parameter $\varepsilon = m/\hat{m} = 1/N$ (where $\hat{m}$ is the mass of the nanosystem and $m$ is the mass of a typical atomic mass) around which the multiscale perturbation expansions is based (Section IV).